\documentclass[conference,a4paper]{IEEEtran}
\IEEEoverridecommandlockouts
\usepackage{cite}
\usepackage{amsmath,amssymb,amsfonts}
\usepackage{algorithmic}
\usepackage{graphicx}
\usepackage{textcomp}
\usepackage{xcolor}
\usepackage{soul}
\usepackage[hidelinks, bookmarks=false]{hyperref}
\usepackage{booktabs}
\usepackage{siunitx}
\usepackage{url}
\usepackage{caption}
\usepackage{subcaption}

\usepackage{standalone}
\usepackage{pgfplots}
\usetikzlibrary{arrows.meta}
\usepackage{tikz}
\usetikzlibrary{positioning,calc}
\pgfplotsset{compat=newest,
    width=11cm,
    height=7cm,
    scale only axis=true,
    max space between ticks=25pt,
    try min ticks=5,
    every axis/.style={
        axis y line=left,
        axis x line=bottom,
        axis line style={thick,->,>=latex, shorten >=-.4cm}
    },
    every axis plot/.append style={thick},
    tick style={black, thick}
}
\tikzset{
    semithick/.style={line width=0.6pt},
}
\usepgfplotslibrary{groupplots}
\usepgfplotslibrary{dateplot}

\newcommand\copyrighttext{%
  \footnotesize \textcopyright  IFIP, 2026. This is the author's version of the work. It is posted here by permission of IFIP for your personal use. Not for redistribution. The definitive version is published in Proceedings of the Wireless On-demand Network systems and Services conference, 2-4 March 2026, Crans-Montana, Switzerland (WONS 2026)}

\newcommand\copyrightnotice{%
\begin{tikzpicture}[remember picture,overlay]
\node[anchor=south,yshift=10pt] at (current page.south) {\fbox{\parbox{\dimexpr0.75\textwidth-\fboxsep-\fboxrule\relax}{\copyrighttext}}};
\end{tikzpicture}%
}

\def\BibTeX{{\rm B\kern-.05em{\sc i\kern-.025em b}\kern-.08em
    T\kern-.1667em\lower.7ex\hbox{E}\kern-.125emX}}

\usepackage[nopostdot,nogroupskip,style=super,nonumberlist,toc]{glossaries}
\newacronym{5gnr}{5G NR}{fifth-generation new radio}
\newacronym{ai}{AI}{artificial intelligence}
\newacronym{lbs}{LBS}{location-based services}
\newacronym{4g}{4G}{fourth-generation}
\newacronym{3gpp}{3GPP}{third generation partnership project}
\newacronym{5g}{5G}{fifth-generation}
\newacronym{6g}{6G}{sixth-generation}
\newacronym{gnb}{gNB}{Next Generation Node B}
\newacronym{enb}{eNB}{Evolved Node B}
\newacronym{ue}{UE}{user equipment}
\newacronym{tdoa}{TDoA}{time difference of arrival}
\newacronym{ul-tdoa}{UL-TDoA}{uplink time difference of arrival}
\newacronym{cn}{CN}{Core Network}
\newacronym{nrppa}{NRPPa}{NR positioning protocol annex}
\newacronym{ul}{UL}{Uplink}
\newacronym{srs}{SRS}{sounding reference signal}
\newacronym{ul-srs}{UL-SRS}{Uplink Sounding Reference Signal}
\newacronym{ofdm}{OFDM}{orthogonal frequency division multiplexing}
\newacronym{csi}{CSI}{channel state information}
\newacronym{oran}{O-RAN}{open radio access network}
\newacronym{oai}{OAI}{OpenAirInterface}
\newacronym{sdrs}{SDRs}{software-defined radios}
\newacronym{snr}{SNR}{signal-to-noise ratio}
\newacronym{cots}{COTS}{commercial off-the-shelf}
\newacronym{cu}{CU}{centralized unit}
\newacronym{du}{DU}{distributed unit}
\newacronym{cu-cp}{CU-CP}{CU control
plane}
\newacronym{cu-up}{CU-UP}{CU user plane}
\newacronym{mac}{MAC}{medium access control}
\newacronym{phy}{PHY}{physical}
\newacronym{ric}{RIC}{RAN intelligent controller}
\newacronym{smo}{SMO}{service management
and orchestration}
\newacronym{TX}{TX}{transmitter}
\newacronym{RX}{RX}{receiver}
\newacronym{ifft}{IFFT}{inverse fast Fourier transform}
\newacronym{idft}{IDFT}{inverse discrete Fourier transform}
\newacronym{ru}{RU}{radio unit}
\newacronym{re}{RE}{resource element}
\newacronym{fr}{FR}{frequency range}
\newacronym{ms}{ms}{milliseconds}
\newacronym{rb}{RB}{resource block}
\newacronym{scs}{SCS}{subcarrier spacing}
\newacronym{rnti}{RNTI}{radio network temporary identifier}
\newacronym{fft}{FFT}{fast Fourier transform}
\newacronym{tdd}{TDD}{time division duplex}
\newacronym{fdd}{FDD}{frequency division duplex}
\newacronym{mimo}{MIMO}{Multiple-Input-Multiple-Output}
\newacronym{dft}{DFT}{discrete Fourier transform}
\newacronym{gpu}{GPU}{Graphics Processing Unit}
\newacronym{ml}{ML}{machine learning}
\newacronym{cir}{CIR}{channel impulse response}
\newacronym{mpc}{MPC}{multipath component}
\newacronym{los}{LoS}{line-of-sight}
\newacronym{nlos}{NLoS}{non-line-of-sight}
\newacronym{cfr}{CFR}{channel frequency response}
\newacronym{ran}{RAN}{radio access network}
\newacronym{near-rt-ric}{Near-RT RIC}{Near real-time RAN Intelligent Controller}
\newacronym{non-rt-ric}{Non-RT RIC}{Non real-time RAN Intelligent Controller}
\newacronym{cc}{CC}{Channel Charting}
\newacronym{l1}{L1}{Layer 1}
\newacronym{l2}{L2}{Layer 2}
\newacronym{scf}{SCF}{Small Cell Forum}
\newacronym{fapi}{FAPI}{Functional Application Platform Interface}
\newacronym{api}{API}{Application Programming Interface}
\newacronym{e2sm}{E2SM}{E2 Service Model}
\newacronym{e2ap}{E2AP}{E2 Application Protocol}
\newacronym{xapp}{xApp}{external application}
\newacronym{trp}{TRP}{Transmission Reception Points}
\newacronym{imu}{IMU}{inertial measurement unit}
\newacronym{mae}{MAE}{mean absolute error}
\newacronym{pso}{PSO}{particle swarm optimization}
\newacronym{ptp}{PTP}{Precision Time Protocol}
\newacronym{lmf}{LMF}{location management function}
\newacronym{http}{HTTP}{Hypertext Transfer Protocol}

\begin{document}
\glsdisablehyper
\title{An O-RAN Framework for AI/ML-Based Localization with OpenAirInterface and FlexRIC}

\author{\IEEEauthorblockN{Nada Bouknana\IEEEauthorrefmark{1}, Mohsen Ahadi\IEEEauthorrefmark{1} , Florian Kaltenberger\IEEEauthorrefmark{1}, Robert Schmidt\IEEEauthorrefmark{2} }
\IEEEauthorblockA{\IEEEauthorrefmark{1}EURECOM, Sophia Antipolis, France\\ Email:\{nada.bouknana,mohsen.ahadi,florian.kaltenberger\}@eurecom.fr}
\IEEEauthorblockA{\IEEEauthorrefmark{2}OpenAirInterface Software Alliance, Sophia Antipolis, France \\ Email: robert.schmidt@openairinterface.org}}

\maketitle
\copyrightnotice
\begin{abstract}
Localization is increasingly becoming an integral component of wireless cellular networks.
The advent of \gls{ai} and \gls{ml} based localization algorithms presents potential for enhancing localization accuracy.
Nevertheless, current standardization efforts in the \gls{3gpp} and the O-RAN Alliance do not support AI/ML-based localization.
In order to close this standardization gap, this paper describes an O-RAN framework that enables the integration of AI/ML-based localization algorithms for real-time deployments and testing.
Specifically, our framework includes an O-RAN \gls{e2sm} and the corresponding \gls{ran} function , which exposes the \gls{ul-srs} channel estimates from the E2 agent to the \gls{near-rt-ric}.
Moreover, our framework includes, as an example, a real-time localization \gls{xapp}, which leverages the custom E2SM-SRS in order to execute continuous inference on a trained \gls{cc} model, which is an emerging self-supervised method for radio-based localization.
Our framework is implemented with \gls{oai} and FlexRIC, democratizing access to AI-driven positioning research and fostering collaboration.
Furthermore, we validate our approach with the \acrshort{cc} \acrshort{xapp} in real-world conditions using an O-RAN based localization testbed at EURECOM.
The results demonstrate the feasibility of our framework in enabling real-time AI/ML localization and show the potential of O-RAN in empowering positioning use cases for next-generation AI-native networks.
\end{abstract}

\begin{IEEEkeywords}
5G, O-RAN, Localization, xApp, Channel Charting, OAI, FlexRIC
\end{IEEEkeywords}

\section{Introduction}
\glsreset{3gpp}
\glsreset{ai}
\glsreset{ml}
\glsreset{ul-srs}
\glsreset{cc}
\glsreset{xapp}
\glsreset{e2sm}
\glsreset{ran}
\glsreset{near-rt-ric}
\glsreset{oai}
Localization continues to evolve as a key technology for current and next-generation networks.
By leveraging large bandwidths, high frequencies, and massive \gls{mimo} capabilities, \gls{5g} systems can enable precise and reliable positioning.
In addition, the service-based architecture and ultra-lean design of 5G make it a cost-effective solution.
Furthermore, 5G's focus on industrial verticals aligns with the requirements of \gls{lbs}.

The \gls{3gpp} positioning algorithms are geometry-based. They rely on channel parameters for position estimation, such as propagation delay, angle, or signal strength\cite{3gpp-TS38.305}. These measurements are then processed by localization algorithms to estimate the position of the \gls{ue} \cite{5gpos_tuto}.
However, these methods rely on line-of-sight (LOS) conditions.
Consequently, their performance can  degrade significantly in \gls{nlos} or dense multipath environments.

While conventional methods are limited in terms of accuracy and performance by the propagation conditions, emerging  \gls{ai} and \gls{ml} methods offer a  promising solution to overcome these limitations, and provide reliable and accurate positioning.
\Gls{csi} contains features that can be exploited for positioning through AI/ML data-driven tools, either to assist traditional methods or to perform direct positioning.
Moreover, AI/ML models for positioning perform well in the presence of \gls{nlos} conditions, in indoor environments with dense multipath propagation, and even when the number of available antennas is limited or the user is in motion\cite{ml_pos_survey}.

The current \gls{3gpp} \gls{5g} positioning architecture does not support the integration of AI/ML algorithms.
However, recent studies in \gls{3gpp} consider the adoption of AI/ML for positioning \cite{3gpp-TR38.843}.
This paves the way to adopting an AI-native architecture for the \gls{6g}.

Building on the principles of openness, virtualization, intelligence, and programmability. \Gls{oran} offers promising prospects for the integration of AI/ML algorithms in the \gls{ran}.
To this end, the O-RAN standardized interfaces and components can offer a flexible and highly programmable solution.
Moreover, the O-RAN Alliance defines a set of requirements for integrating AI/ML in the O-RAN architecture \cite{oran-wg2-ai}.
The \gls{non-rt-ric} can be used for training AI/ML models and providing updated models to the \gls{near-rt-ric}. Furthermore, the \gls{near-rt-ric} can host \glspl{xapp} to perform the AI/ML inference.
This approach can also be applied to positioning, which is considered in the O-RAN Alliance Use Cases Detailed Specification\cite{oran-wg1-specific-usecases}.
Nonetheless, the O-RAN Alliance does not specify the detailed message flow, or any dedicated \gls{e2sm} for enabling AI/ML-based positioning.

In order to address this standardization gap, this paper proposes and validates an O-RAN framework which enables AI/ML-based localization for real-time deployments.
Our proposed solution leverages the O-RAN E2 interface in order to transport the channel estimates of the \gls{ul-srs} to the \gls{near-rt-ric} and enable real-time AI/ML localization inference in an \gls{xapp}.
We provide a concrete open-source implementation of a new custom O-RAN compatible E2SM using FlexRIC~\cite{flexric-paper} and integrate it with \gls{oai}~\cite{kaltenberger-oai25}.
Furthermore, we provide, as an example, a real-time localization \gls{xapp} that performs \gls{cc} inference, which is an emerging self-supervised data-driven method for \gls{ue} localization in wireless networks.
It is based on learning the similarities between high dimensional \gls{csi} measurements and applying dimensionality reduction techniques to the \gls{csi} to produce a lower dimensional embedding called channel chart.
The produced channel chart contains information about the user location\cite{cc_first}.
The xApp uses a pre-trained \gls{cc} model based on a novel algorithm that we previously proposed in \cite{ahadi2025tdoabasedselfsupervisedchannelcharting}.
This paper also discusses experimental results obtained from real-time testing under real-world conditions with an O-RAN based localization testbed at EURECOM.

Several works have studied and evaluated wireless localization techniques in 5G using \gls{oai}.
For instance, authors in \cite{oai_pos_tools} provide a detailed guide on the implementation of positioning features in OAI, mostly focusing on the \gls{phy} layer and on the useful tools for data extraction which is necessary for post-processing and analysis.
Authors in \cite{11152876} demonstrated the integration of \gls{nrppa} and Uplink \gls{tdoa} using the \gls{ul-srs} for positioning in OAI.
Finally, authors in \cite{ahadi2025experimental} presented experimental results of 5G positioning in real-conditions using experimental testbeds that are integrated with OAI, they also demonstrate a new framework suitable for AI/ML-based positioning in beyond-5G. However, the implementation does not use O-RAN standardized protocols.
In summary, the primary focus of the previous works was implementing and benchmarking different positioning algorithms with OAI. However, and to the best of the authors' knowledge, our work is the first to propose a framework, with a concrete implementation, which enables AI/ML localization inference using FlexRIC and OAI.

The main contributions of this paper can be summarized as follows: we present an O-RAN-based framework, which constitutes of a custom O-RAN E2SM and an xApp, that enable real-time AI/ML inference for SRS positioning. We provide a concrete open-source implementation, of the E2SM and a specialized CC localization xApp, using OAI and FlexRIC.
We validate our solution in real-world conditions using an O-RAN-based localization testbed at EURECOM.

The remainder of this paper is organized as follows.
Section \ref{sec2} describes the system model.
Section \ref{sec3} presents the deployed CC model.
Section \ref{sec4} discusses the implementation details of the custom O-RAN service model, which transports the UL-SRS channel estimates from the E2 agent to the Near-RT RIC, it also describes the implementation of the real-time CC-based localization xApp.
Section \ref{sec5} describes the experimental validation conducted using the Firecell GEO5G testbed at EURECOM, and analyzes the obtained results.
Finally, section \ref{sec7} concludes the paper and provides future perspectives.

\section{System Model} \label{sec2}
\begin{figure}[t]
    \centering
     \includegraphics[width=.5\textwidth]{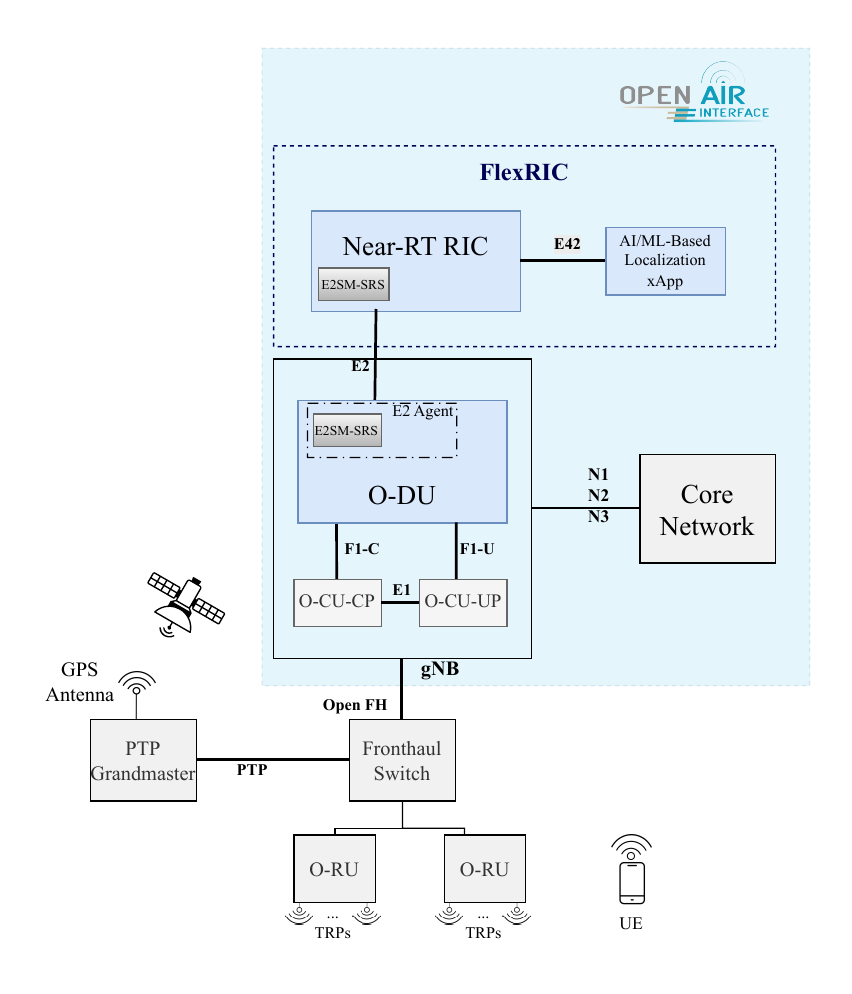}
     \caption{AI/ML-based localization system model}
     \label{fig:ric-archi}
\end{figure}
As shown in Figure~\ref{fig:ric-archi}, our AI/ML-based 5G localization system model consists of an O-RAN compliant \gls{5g} system, with the \gls{cn} and the \gls{ran} which includes the \gls{gnb} and the \gls{ue}.
The UE transmits the UL-SRS signal to one or multiple gNBs.
While we consider one gNB in this work, the approach can be extended to mutiple gNBs.
The gNB is further disaggregated into \gls{cu}, \gls{du}, and \gls{ru} (O-CU, O-DU, and O-RU following the O-RAN architecture).
The O-DU and O-RU communicate via the O-RAN 7.2 fronthaul interface.

There are $K$ RUs each with known \gls{trp} locations $\mathbf{x}_{m_k}$ where $m_k\in \{1,\dots,M_k\}$.
Every TRP receives the transmitted UL-SRS signal.
The O-DU estimates the UL-SRS channel.
The estimated \gls{cfr} of the link between the $m_k$-th \gls{trp} 
of the $k$-th RU and the UE at time step $t$, spanning all \gls{ofdm} $N_{\text{fft}}$ sub-carriers, is denoted as 
$\mathbf{w}_{k,m_k,t} \in \mathbb{C}^{N_{\text{fft}}}$.
The E2 interface transports the estimated frequency-domain \gls{mimo} channel matrix from the E2 agent (O-DU in our system model) to the Near-RT RIC.
An xApp, which is part of the RIC, continuously performs the machine learning inference and updates the UE position.

We note the \gls{cir} by  
$\mathbf{h}_{k,m_k,t} \in \mathbb{C}^{N_{\text{fft}}}$ 
which is obtained by applying an \gls{idft} to the \gls{cfr}.
In our configuration, the \gls{idft} is performed by the \gls{xapp}.
Collecting the CIRs from all $M_k$ TRPs of RU $k$ yields the RU-level CIR matrix:
\begin{equation}
\mathbf{H}_{k,t} = 
\begin{bmatrix} 
\mathbf{h}_{k,1,t} \\ 
\mathbf{h}_{k,2,t} \\ 
\vdots \\ 
\mathbf{h}_{k,M_k,t}
\end{bmatrix}
\in \mathbb{C}^{M_k \times N_{\text{fft}}}
\end{equation}

Similarly, the global CIR matrix across all $K$ RUs is
\begin{equation}
  \mathbf{H}_{t} = 
  \begin{bmatrix} 
  \mathbf{H}_{1,t} \\ 
  \vdots \\ 
  \mathbf{H}_{K,t}
  \end{bmatrix}
  \in \mathbb{C}^{M \times N_{\text{fft}}}
\end{equation}
where $M = \sum_{k=1}^K M_k$ is the total number of TRPs across all RUs. 
The global CIR $\mathbf{H}_{t}$ is subsequently processed for ML inference.

\section{Channel Charting} \label{sec3}
In this section, we provide a brief overview of the deployed CC model. The development of the CC algorithm is not part of this work, and we refer the reader to our previous work which proposed this novel CC algorithm \cite{ahadi2025tdoabasedselfsupervisedchannelcharting}.

We aim to learn a self-supervised CC function 
$f_{\boldsymbol{\theta}}(\cdot)$ that maps a high dimensional channel measurement into a two-dimensional embedding space:
\[
f_{\boldsymbol{\theta}} : \mathbb{R}^{M \times C} \rightarrow \mathbb{R}^{2}.
\]

The objective of this mapping is to preserve the inherent similarities of the radio 
channel domain within the embedded space. For example, channel similarities caused by nearby measurement locations in the physical space should also appear close to each other in the embedded space.
The estimation of the CC function is carried out by training a Convolutional Neural Network (CNN) model on a 
dataset of channel measurements that have been cleaned and pre-processed to extract 
relevant features. The training yields an optimized mapping function 
$f_{\boldsymbol{\theta}}^{\ast}$, which is then used to predict the UE position from 
unseen CIR data after applying the same pre-processing steps.

\subsection{Data Pre-Processing}
\begin{figure}[t]
    \centering
     \includegraphics[width=.5\textwidth]{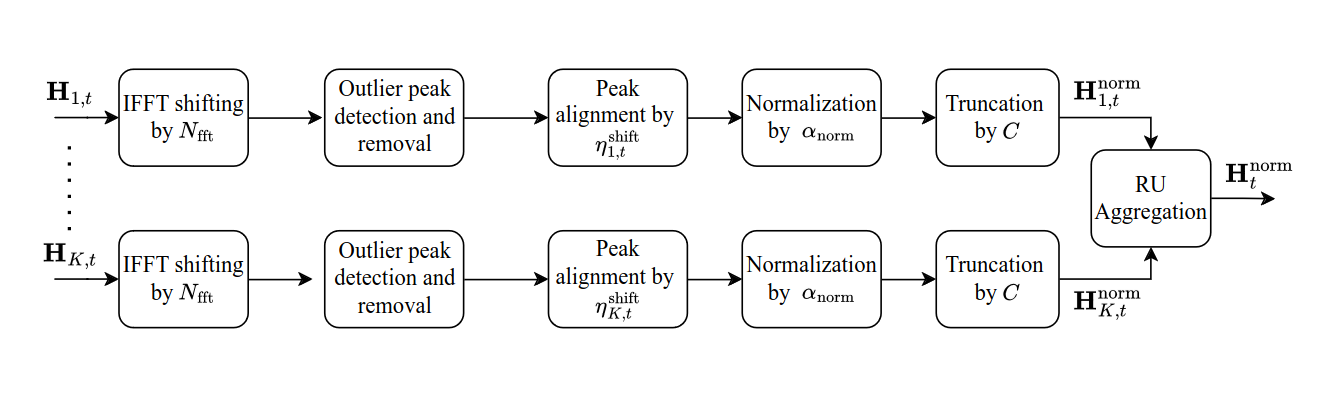}
     \caption{CC pre-processing pipeline \protect \cite{ahadi2025tdoabasedselfsupervisedchannelcharting}}
     \label{fig:cc-preproc}
\end{figure}

Figure~\ref{fig:cc-preproc} show the pre-processing steps of the CC model.
The first step is to apply \gls{ifft} shifting to position the zero-frequency component at the center of the time-domain CIR output.
Then, we remove outliers, which are the result of synchronization hardware impairments present in the testbed.
All of this results in a \gls{tdoa}-aligned RU-level CIR matrix \({\mathbf{H}}_{k,t}^{\text{shifted}}\).
We normalize \({\mathbf{H}}_{k,t}^{\text{shifted}}\) by the normalization factor \(\alpha_{\text{norm}} = \max\limits_{k,m_k,t} ({ |\mathbf{h}}_{k,m_k,t}^{\text{shifted}}|)\), which is computed from the training data and reused during testing to ensure consistency.
Finally, we truncate the result to only contain the first $C$ \gls{fft} indices.
The details about each step are explained in \cite{ahadi2025tdoabasedselfsupervisedchannelcharting}.
The final result is a normalized and truncated channel matrix, expressed as:
\begin{equation}\label{eq:normalized}
\mathbf{H}_{t}^{\text{norm}} 
= \frac{1}{\alpha_{\text{norm}}}
\begin{bmatrix}
|\mathbf{h}_{1,t}^{\text{shifted}}| \\[4pt] 
|\mathbf{h}_{2,t}^{\text{shifted}}| \\[2pt] 
\vdots \\[2pt] 
|\mathbf{h}_{M,t}^{\text{shifted}}|
\end{bmatrix}
\in \mathbb{R}^{M \times C}
\end{equation}
\subsection{Training Phase}
After collecting a substantial set of channel measurements under diverse conditions and trajectories, the data are pre-processed based on the principles from the previous section. The resulting dataset is then prepared and structured for use in training a CNN model, enabling effective feature learning and robust performance.

\begin{figure}[t]
    \centering
    \includegraphics[width=0.35\textwidth]{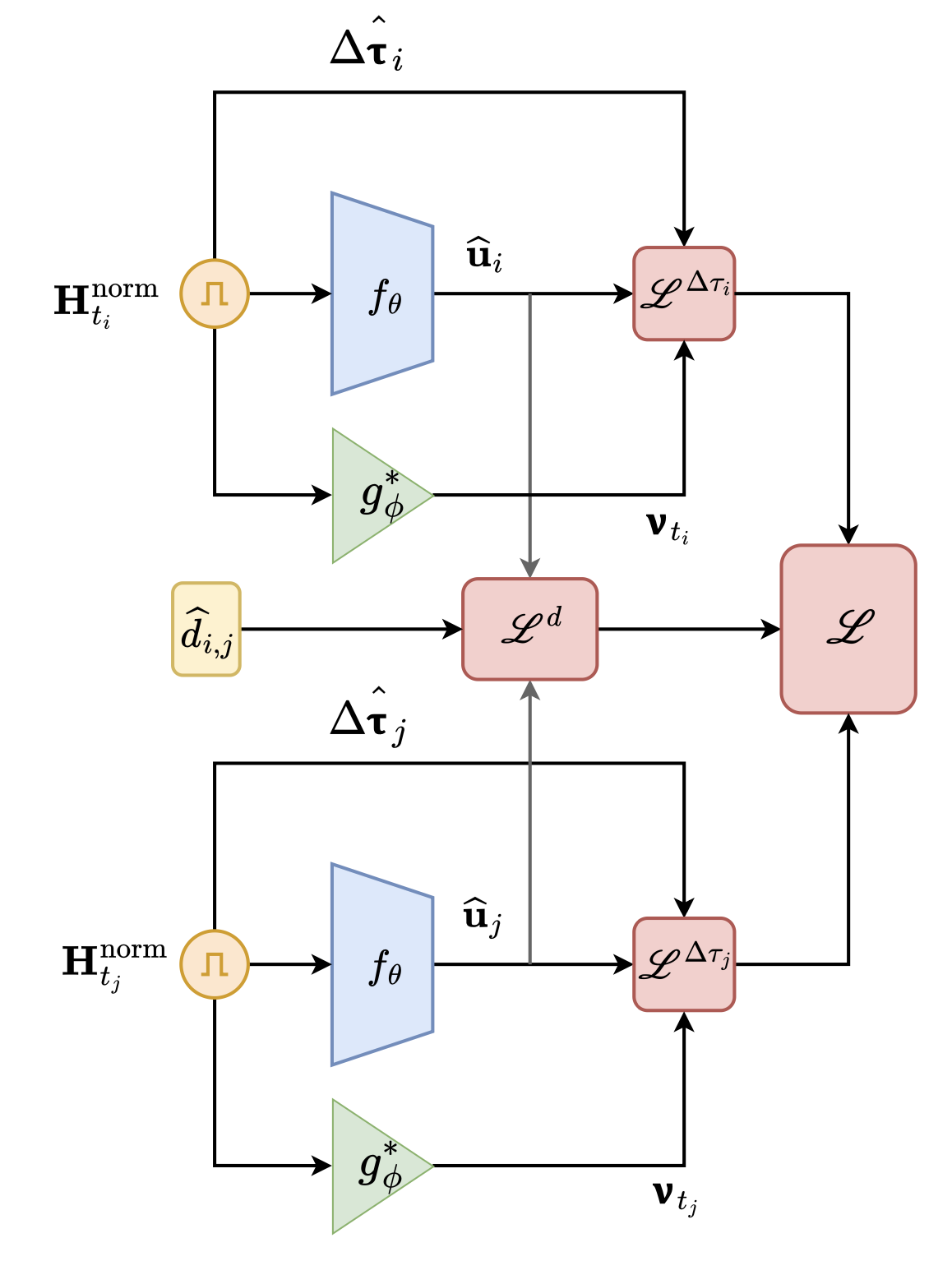}
    \caption{CC Training with CIR and TDoA+displacement \protect \cite{ahadi2025tdoabasedselfsupervisedchannelcharting}}
    \label{fig:cc_training}
\end{figure}

The model training steps are depicted in Figure~\ref{fig:cc_training}.
To train the model, we proceed as described in \cite{ahadi2025tdoabasedselfsupervisedchannelcharting}, we begin by randomly selecting pairs of CIR samples $(\mathbf{H}^{\text{norm}}_{t_i}, \mathbf{H}^{\text{norm}}_{t_j})$ as well as their corresponding TDoAs $(\Delta\hat{\tau}_{t_i},  \Delta\hat{\tau}_{t_j})$ from the dataset, with timestamps $t_i$ and $t_j$ if their temporal interval satisfies $|t_j - t_i| \leq \epsilon$. Otherwise, they are discarded, and new pairs are drawn.

We consider two loss functions, the first loss function $\ell^{\Delta\tau}_{t_i,t_j}$ reflects the TDoA loss, which discards the NLoS-corrupted measurements using binary masking, where a classification function, denoted as $g_{\boldsymbol{\phi}}^{*}$, maps normalized CIR measurements to a binary LoS/NLoS masking vector.
The second loss function $\ell^{d}_{t_i, t_j}$ incorporates the displacement of the UE $\hat{d}_{i,j}$ between timestamps $t_i$ and $t_j$. The displacement is assumed to be available (e.g., measured from an external sensor such as laser or \gls{imu}). In addition, the time interval $\epsilon$ is chosen to avoid sensor noise and bias from \gls{imu}.

While the \gls{tdoa} loss function could embed the \gls{cir} into two dimensions with a global scale, the purpose of displacement loss is to smoothen the fluctuations in a moving scenario.
Thus, the overall training objective integrates two terms across all $T$ time steps, $K$ RUs and their $M_k$ TRPs, can be expressed as:
\begin{align}\label{eq:tdoa_disp_loss}
    \mathcal{L} &=
    \frac{1}{T K (M_k-1)} 
      \sum_{\substack{t_i,t_j=1 \\ |t_i - t_j| < \epsilon}}^{T} 
      \sum_{k=1}^{K} 
      \sum_{\substack{m_k=1 \\ m_k \neq m_{\text{ref}_k}}}^{M_k} \ell^{\Delta\tau}_{t_i,t_j} + \beta \, \ell^{d}_{t_i, t_j},
\end{align}
where the first term $\ell^{\Delta\tau}_{t_i,t_j}$ is the pair-wise TDoA measurement loss, the second term $\ell^{d}_{t_i, t_j}$ is the displacement measurement loss, and $\beta$ controls the relative weight of the displacement term.

\subsection{Testing Phase}
\begin{figure}[t]
    \centering
    \includegraphics[width=0.35\textwidth]{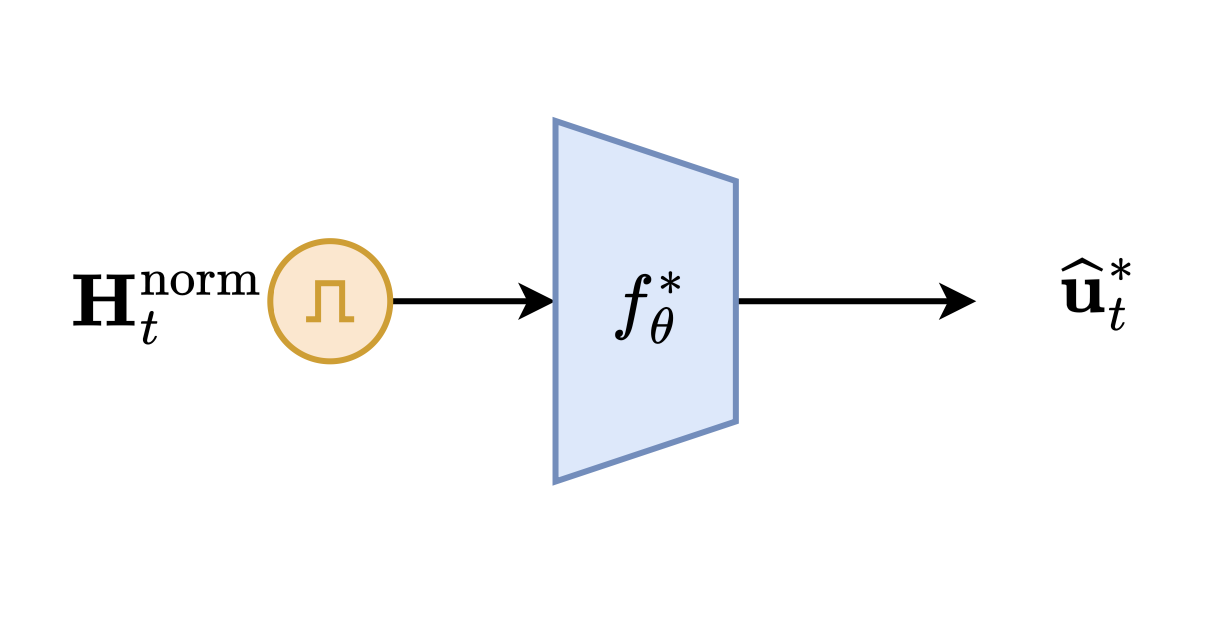}
    \caption{CC Testing \protect \cite{ahadi2025tdoabasedselfsupervisedchannelcharting}}
    \label{fig:cc_testing}
\end{figure}

During training, our model learns to jointly capture both TDoA and displacement information in a mixed LoS/NLoS scenario. 
It is important to note that the testing data are entirely unseen and have not been used during training. 
Consequently, in the testing phase, as shown in Figure~\ref{fig:cc_testing}, the optimized mapping function $f^{\ast}_{\boldsymbol{\theta}}(\cdot)$ operates solely on the pre-processed CIR inputs, without requiring explicit displacement, TDoA feature annotation or NLoS masking.

\section{Implementation} \label{sec4}
In this section, we introduce and detail the implementation of the custom O-RAN service model E2SM-SRS and define the "SRS Positioning" RAN function, which supports the RIC REPORT service, which is used to expose the SRS frequency-domain channel estimates inside the RIC indication message via the E2 interface.
In this work, the service model will be leveraged by a localization xApp that exploits and processes the RIC indication message in order to perform continuous inference on a pre-trained CC model.
The E2SM-SRS and xApps are implemented in FlexRIC and integrated with the E2 agent in OAI.
All the code is available in the \emph{srs\_sm}~\cite{srs-sm} branch in FlexRIC and \emph{srs\_e2}~\cite{srs-e2} branch in \gls{oai}. 
\footnote{If the branches do not exist anymore, use the \emph{dev} branch in FlexRIC and the \emph{develop} branch in OAI}

\subsection{OpenAirInterface Overview}
OpenAirInterface is an open-source project that provides a software-defined implementation of 4G and 5G 3GPP systems.
It is O-RAN compliant as well.
OAI includes the CN, the RAN and the UE.
The OAI platform consists of several coordinated projects, including the RAN, CN, Continuous Integration/Continuous Delivery (CI/CD) pipelines, and Operations, Administration and Maintenance (OAM).
The platform is maintained by the OpenAirInterface Software Alliance (OSA) which defines roadmaps of each project, promotes the software and supports the community.

OAI supports different functional splits such as the F1 split between the CU and the DU and the E1 split between the \gls{cu-up} and the \gls{cu-cp}.
It also supports the O-RAN 7.2 fronthaul split as well as the eCPRI split 8, in order to work with third-party \gls{sdrs} or \glspl{ru}.

OAI also supports the \gls{fapi} \cite{fapi}, which is an interface specified by the \gls{scf} that shows the interplay between \gls{l1} and \gls{l2}.
It allows in-line acceleration with the integration with hardware-accelerated or \gls{gpu} based physical layer implementations, such as NVIDIA’s Aerial L1 stack \cite{x5g_24} \cite{x5g_25}.

Additionally, OAI supports the E2 interface, which allows the OAI gNB, CU, or DU to act as an E2 agent for integration with a Near-RT RIC, in accordance with O-RAN architecture. This enables real-time monitoring, control, and optimization of the RAN through xApps. 
For more details on OAI, we refer the reader to \cite{kaltenberger-oai25}.

\subsection{FlexRIC}

FlexRIC is a software development kit (SDK) that provides an O-RAN compliant Near-RT RIC and xApp SDK.
It consists of server and agent library.
It was designed to follow the 5G principles of ultra-lean design and it is implemented with zero-overhead principle \cite{flexric-paper}.
FlexRIC supports different versions of the \gls{e2ap} (v1.0/2.0/3.0) and implements some O-RAN specified Service Models; the Key Performance Measurement (KPM) and Radio Control (RC) service models.
It also allows the implementation of some custom service models including those for the Medium Access Control (MAC), Packet Data Convergence Protocol (PDCP), and Radio Link Control (RLC) layers.
The xApps can be developed in different high-level programming languages such as C, C++, and Python, offering flexibility and rapid development.

\subsection{E2SM-SRS}
\begin{figure}[t]
      \centering
     \includegraphics[width=.5\textwidth]{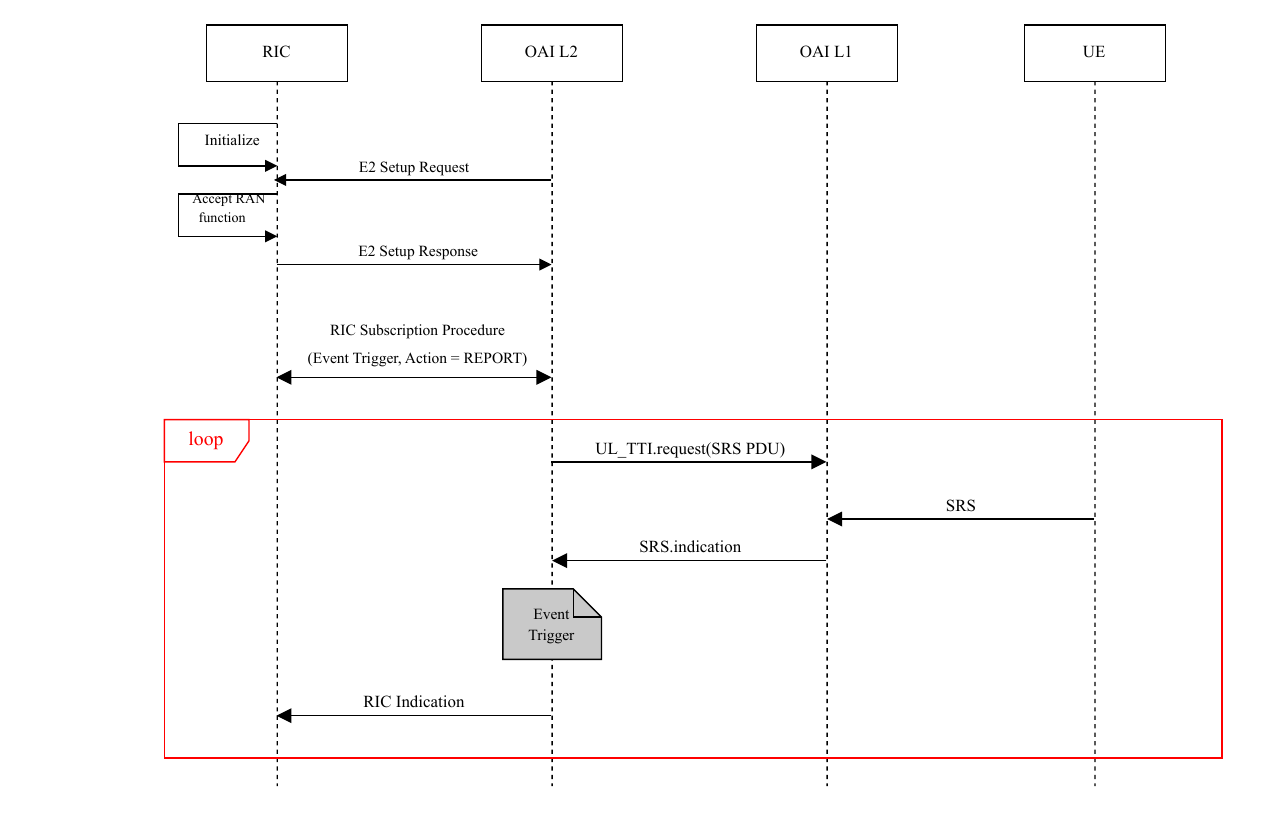}
     \caption{Message flow between the RIC and the RAN entities}
      \label{fig:ric-ind}
\end{figure}

The custom E2SM-SRS service model provides the semantic description of specific fields which contain the data to be transported over the E2 interface.
In our implementation, we leverage the \gls{fapi} SRS procedure from \gls{scf} \cite{fapi} and extend it to the RIC, because the exchanged messages in this procedure contain the SRS channel estimates.
As specified by \gls{scf}, the FAPI SRS.indication message contains fields related to the SRS including a sub-sampled version of the SRS frequency domain channel estimates.
For our purposes, the E2SM-SRS exposes a modified version of the SRS.indication message, which instead contains the channel estimates with the full resolution.
In total, the RIC indication message contains the SRS.indication message and a UE identifier to distinguish between multiple UEs.
Figure~\ref{fig:ric-ind} illustrates the message flow between the RIC and the other RAN entities using the E2SM-SRS.
The RIC establishes connection with the E2 agent (monolithic gNB or DU in this case) via the E2 Setup procedure.
To enable the FAPI SRS procedure, the L2 should include a SRS PDU in UL{\_}TTI.request.
After receiving the SRS from the UE, the L1 will process and forward the SRS response in the SRS.indication message.
The reception of the SRS.indication message by the L2 constitutes the event trigger for our RIC Report Service.
Whenever this trigger is satisfied, the RIC indication message gets forwarded to the Near-RT RIC via the E2 interface.
\subsection{Localization xApp}
In our framework, the Near-RT RIC hosts an xApp that performs localization inference.
As an example, we implemented a specialized localization xApp in C++, that leverages our custom E2SM-SRS and performs continuous inference on a pre-trained CC model.
The xApp is used to evaluate the CC model using the Firecell GEO5G testbed at EURECOM. 
After the xApp receives the RIC indication message, it unpacks the fields of the SRS.indication message.
The next step is the pre-processing of the SRS channel in order to prepare it for ML-inference.
We implemented the steps described in \ref{sec3} using \emph{libtorch}.
Then, the xApp performs the inference using a trained model.
Moreover, the CC predictions are enhanced with a moving average filter.
Furthermore, the xApp integrates a demonstrator, in the form of a real-time graphical user interface(GUI) that plots the SRS channel and the position estimates in a two-dimensional representation of the testbed map.

\subsection{E2 Agent emulator support}
For prototyping and testing purposes, we extended the O-RAN compliant E2 agent emulators provided by FlexRIC to support our new E2SM-SRS for AI/ML-based localization.
In our implementation, the E2 emulator reads CSI samples from the EURECOM 5G SRS CIR dataset~\cite{t8ya-z141-25} and forwards them to the Near-RT RIC.
This can easily be configured to use any other dataset.
This setup enables testing new algorithms within an O-RAN-compatible framework, providing a useful intermediary step before deploying a full end-to-end system.
For reproducibility, we provide a detailed tutorial in \cite{xapp-tuto}.

\section{Experimental Evaluations} \label{sec5}

\subsection{Setup}

\begin{figure}[t]
    \centering
    \includegraphics[width=0.45\textwidth]{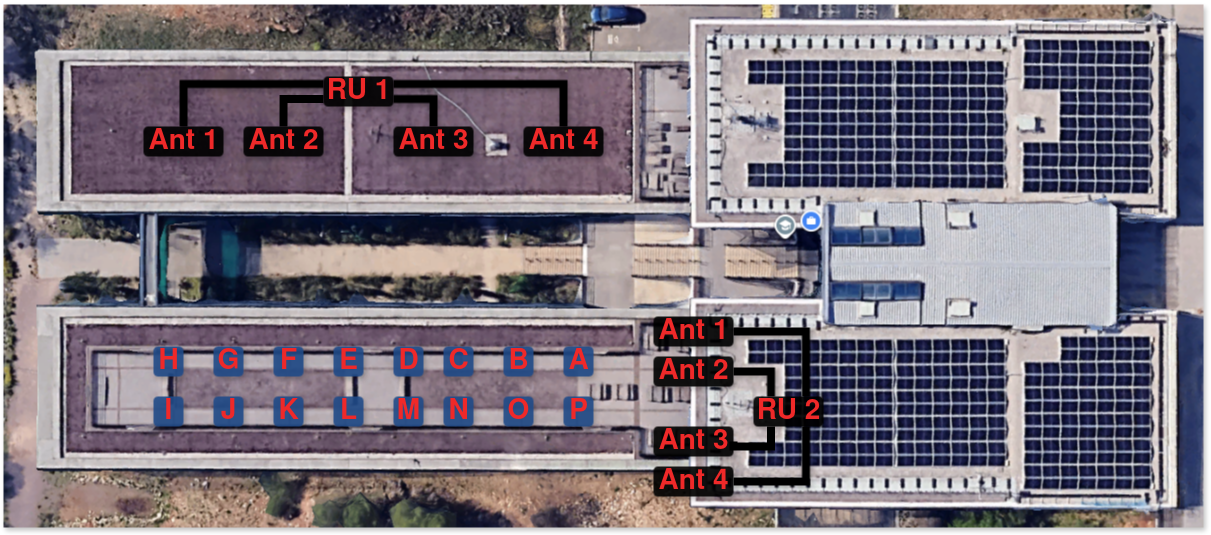}
    \caption{Firecell GEO-5G testbed at EURECOM: Deployment of 2 O-RUs each with 4 distributed TRPs on the south terrace and the north rooftop of EURECOM, with a testing area of size $50\times10$ on the north terrace}
    \label{fig:testbedmap}
\end{figure}
To validate the proposed framework, we conducted experiments in real-world conditions using the Firecell GEO-5G testbed at EURECOM.
The testbed is built on top of EURECOM's OpenXG platform, which provides high-speed fiber-connected computing and switching infrastructure. 
This environment supports virtual 5G deployments with USRPs and O-RAN radios, and integrates OAI for running virtual network functions.  

Thanks to our recent contributions to OAI RAN~\cite{11152876}, the testbed can flexibly operate under either a single-gNB with multiple RUs, or a multi-gNB with multiple RUs architecture.
To evaluate the new localization features in OAI, we deployed two VVDN O-RAN RUs~\cite{vvdn_oran_portfolio_2023}, provided by Firecell~\cite{firecell_5g}, configured with a single gNB (CU-DU) and integrated into the OpenXG infrastructure.

Each RU is equipped with four distributed Panorama directional antennas~\cite{panorama_antennas}, mounted on the roof railings and connected through low-loss cables. This results in 
a total of eight TRPs, providing coverage over a $50 \times 10$~m testing area on the north terrace of the EURECOM building (see Figure~\ref{fig:testbedmap}).
There are 16 test points (A to P) distributed in the testing area. All test points accurate locations in a local Cartesian coordinate are calculated using true distances between them and the antennas by a laser ranging tool. The reference for our local coordinate is TRP 1 on RU1.

In our configuration, we use 5G New Radio (5G NR) band n77 with $100$ Mhz maximum bandwidth, in time division duplex (TDD) mode using $30$ kHz subcarrier spacing.
The CU and DU run on the same server.
The CN runs on a separate server in a docker environment.
FlexRIC, which includes, the Near-RT RIC and the xApp (with the GUI) runs on bare metal on a separate machine connected to the network.
Deploying FlexRIC in a docker environment is also possible.

\subsection{Results}
This section presents the results for the performance evaluation of our O-RAN-based localization framework.
We assess the accuracy of \gls{cc}, followed by an analysis of the latency. The results are compared with the 3GPP \gls{ul-tdoa} positioning.
\subsubsection{Accuracy}

Two scenarios are considered: one for a static UE and one for a moving UE.
The first experimental scenario involves placing one commercial off-the-shelf (COTS) UE on each test point with a fixed height using a tripod. The xApp is then executed to perform the CC inference. We store the CC predictions on a \emph{csv} file in real-time. The Euclidean distance between the CC predictions and the ground truth measurements represents the error in the measurements.
We consider the 90th percentile of the error and the \gls{mae} as performance metrics of the positioning accuracy.

In the moving scenario, a handheld UE moves in a random trajectory along the test points, and the CC estimations are illustrated in real-time over the GUI. This provides valuable qualitative insights into the model’s behavior and allows for observation of how changes in the wireless environment affect the CC predictions.

\begin{figure}[htbp]
  \centering
    \resizebox{\linewidth}{!}{\input{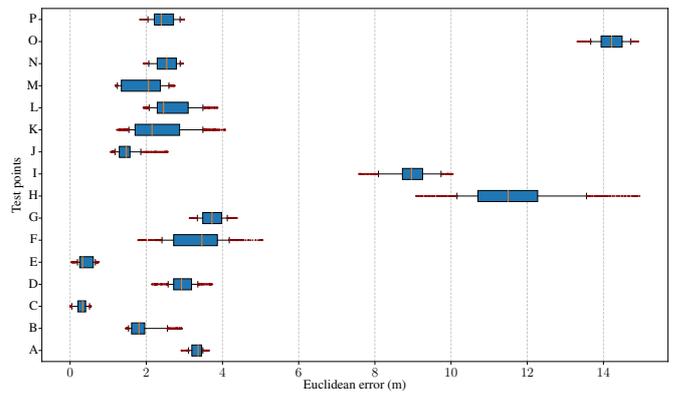}}
    \caption{Box plots of Euclidean error of all test points, where the whiskers are set to the 10th and 90th percentile, the orange line shows the median and the red dots represent the outliers.}
    \label{fig:box_plot}
\end{figure}

The box plots in Figure~\ref{fig:box_plot} visualize the distributions of the error in all 16 test points.
The lower whisker is set at the 10th percentile, and the higher whisker is set at the 90th percentile.
We observe different error distributions across the test points with varying skewness.
For most test points, the median error, represented by the orange line, lies in the expected range under 4~m in most of the test points except test points H,I and O.

\begin{figure}
     \hspace*{-0.5cm}
       \resizebox{\linewidth}{!}{\input{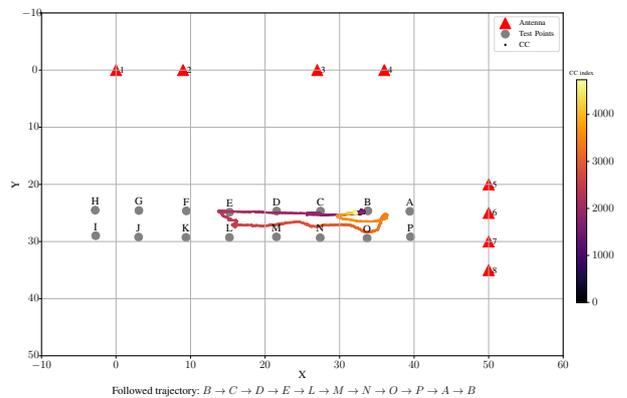}}
     \caption{CC UE tracking in a moving scenario}
     \label{fig:ue_trajectory_results}
\end{figure}

Figure~\ref{fig:ue_trajectory_results} shows the CC predictions in a moving scenario for a UE taking the following trajectory: $ B \rightarrow C \rightarrow D \rightarrow E \rightarrow L \rightarrow M \rightarrow N \rightarrow O \rightarrow P \rightarrow A \rightarrow B$.
For error analysis in this moving scenario, existing tools such as those based on real-time kinematic positioning (RTK) can deliver cm-level positioning accuracy which can be considered as ground truth for outdoor deployments.
However, it remains complicated to evaluate the positioning error.
This is mainly due to practical impairments, such as synchronizing the CC predictions with RTK measurements.
In future developments, more reliable and robust methods will be investigated to assess the positioning accuracy for a moving UE.

We also compare the accuracy of \gls{cc} with a classical 3GPP \gls{ul-tdoa} protocol employing \gls{pso} algorithm to compute the position~\cite{11152876}.
Table~\ref{tab:mae_comparison} shows the comparison of the \gls{mae} for the same 16 test points.
It can be observed that \gls{ul-tdoa}  performs  better for most of the test points, but it should be noted that the measurements were done at different times and it could be that the environment has changed in between. This highlights the very important fact that neural networks for localization are very site and environment specific and should be regularly retrained.

\begin{table}[ht]
\caption{Comparison of the MAE (in meters) for the test points}
\label{tab:mae_comparison}
\centering
\sisetup{
  table-number-alignment = center,
  round-mode = places,
  round-precision = 2
}
\begin{tabular}{c S S}
\toprule
Test point & {MAE$_\text{CC}$} & {MAE$_\text{UL-TDOA}$} \\
\midrule
A & 3.33  & 1.88 \\
B & 1.87  & 1.13 \\
C & 0.30  & 0.82 \\
D & 2.94  & 0.83 \\
E & 0.40  & 0.95 \\
F & 3.35  & 0.55 \\
G & 3.73  & 0.92 \\
H & 11.64 & 0.56 \\
I & 8.95  & 0.60 \\
J & 1.49  & 0.83 \\
K & 2.33  & 0.82 \\
L & 2.65  & 1.20 \\
M & 1.94  & 0.64 \\
N & 2.51  & 0.68 \\
O & 14.19 & 0.97 \\
P & 2.43  & 1.98 \\
\bottomrule
\end{tabular}
\end{table}

\subsubsection{Latency}
Finally, we evaluate the latency of the proposed O-RAN framework and compare it with that of the 3GPP \gls{nrppa}-based positioning architecture~\cite{11152876}.
Latency is measured over multiple trails by recording timestamps at the relevant entities.
In the O-RAN framework, timestamps are recorded at the E2 agent, and xApp.
For the \gls{3gpp} architecture, timestamps are recorded at the \gls{lmf} and the overall latency is measured from the initiation of location request until the reception of the positioning result.
In our experimental setup, all the involved servers are synchronized using \gls{ptp} in order to ensure timing consistency.

As explained in Section \ref{sec4}, the RIC indication procedure is event-driven.
Thus, the end-to-end latency is defined as the elapsed time between the event trigger —corresponding to the reception of a FAPI SRS indication message— and the delivery of the \gls{cc} prediction produced by the xApp.

To provide a detailed analysis, we decompose the overall latency into two components.
First, the RIC indication message latency is measured as the time elapsed from when the E2 agent generates and sends the RIC indication message, until its reception by the \gls{xapp}.
Second, the xApp processing latency is measured separately, capturing the unpacking of the SRS.indication message, the preprocessing, and the CC inference.
The end-to-end latency corresponds to the sum of these two components.
In addition, the inference latency is measured in isolation to quantify its contribution to the total end-to-end latency.

Experimental results show an average end-to-end latency of 7.749~ms, out of which 4.914~ms corresponds the processing latency.
The \gls{cc} inference latency is 4.230~ms, accounting for around 86\% of the processing latency.

For comparison, we also measured the latency of the 3GPP \gls{ul-tdoa} architecture where the \gls{lmf} is using \gls{pso} as positioning algorithm~\cite{11152876}.
Unlike the O-RAN architecture, the \gls{3gpp} positioning protocol is based on request and response: the positioning process begins when an external \gls{api} sends a location request of a UE to the \gls{lmf}.
In the implementation, this is issued via a single-line \texttt{curl} command, which sends an \gls{http} POST request to the determine location \gls{api}.
After computing the UE location, the \gls{lmf} returns a response.

The overall latency is measured by collecting timing information provided by the \texttt{curl} command.
As with the RIC approach, we also separately measure the latency of the \gls{pso} algorithm.
However, it is important to note that conducting a fair comparison between the two approaches remains challenging.
A substantial portion of the latency in the 3GPP architecture is spent waiting for the SRS measurement response.
Consequently, the measurements obtained via the \texttt{curl} command represent an upper bound on the end-to-end latency.

The results of the 3GPP NRPPa-based approach show that the average overall latency is 32.125~ms while the average latency of the PSO is 0.240~ms.
While the NRPPa protocol latency is substantially higher than that observed in the RIC procedure, the \gls{pso} computation is significantly lower than the \gls{cc} inference time in the xApp.
However, it should be noted that we did not seek to optimize the performance of CC and there is room for improvement  by using a different implementation on a more efficient inference engine, or GPU acceleration.

\section{Conclusions} \label{sec7}
This paper provided an O-RAN based framework for the integration of real-time AI/ML-based localization inference in 3GPP and O-RAN compliant 5G systems and beyond.
The implementation is fully open-source, based on OAI and FlexRIC, provides a new custom E2SM that successfully transports the UL-SRS channel estimates to the NearRT-RIC via the E2 interface.
Localization is performed in an xApp hosted in the Near-RT RIC.
The approach was integrated with a CC xApp and validated in real-world conditions using the Firecell GEO5G testbed at EURECOM. 

The purpose of the paper was to show how the O-RAN architecture can be leveraged to perform AI/ML-based positioning. The algorithms and implementations of the xApp used in this paper is just an example and not the main contribution of this work. 

As future work, we plan on extending this framework to multi-cell deployments, integrating other ML methods, as well as extending to sensing.
We also consider integrating the Near RT-RIC with a Non RT-RIC to perform the AI/ML model training.

\bibliographystyle{IEEEtran}
\bibliography{IEEEabrv,refs.bib}

\end{document}